\documentclass[twocolumn,showpacs,preprintnumbers,amsmath,amssymb]{revtex4}
\usepackage{graphicx}
\usepackage{color}
\begin{document}

\title{Stress-Energy Tensor Induced by Bulk Dirac Spinor in Randall-Sundrum Model}

\author{Shu-Heng Shao$^{1,3}$}\email{b95202055@ntu.edu.tw}

\author{Pisin Chen$^{1,2,3,4}$}\email{pisinchen@phys.ntu.edu.tw}
\author{Je-An Gu$^3$}\email{jagu@ntu.edu.tw}
\affiliation{1. Department of Physics, 
National Taiwan University, Taipei 10617, Taiwan, R.O.C.\\
2. Graduate Institute of Astrophysics, 
National Taiwan University, Taipei 10617, Taiwan, R.O.C.\\
3. Leung Center for Cosmology and Particle Astrophysics, National Taiwan University, Taipei 10617, Taiwan, R.O.C.\\
4. Kavli Institute for Particle Astrophysics and Cosmology,
SLAC National Accelerator Laboratory, Menlo Park, CA 94025, U.S.A.}

\date{\today}
\begin{abstract}
Motivated by the possible extension into a supersymmetric Randall-Sundrum (RS) model, we investigate the properties of the vacuum expectation value (VEV) of the stress-energy tensor for a quantized bulk Dirac spinor field in the RS geometry and compare it with that for a real scalar field. This is carried out via the Green function method based on first principles without invoking the degeneracy factor, whose validity in a warp geometry is {\it a priori} unassured. In addition, we investigate the local behavior of the Casimir energy near the two branes. One salient feature we found is that the surface divergences near the two branes have opposite signs. We argue that this is a generic feature of the fermionic Casimir energy density due to its parity transformation in the fifth dimension. Furthermore, we investigate the self-consistency of the RS metric under the quantum correction due to the stress-energy tensor. It is shown that the VEV of the stress-energy tensor and the classical one become comparable near the visible brane if $k\simeq M\simeq M_{Pl}$ (the requirement of no hierarchy problem), where $k$ is the curvature of the RS warped geometry and $M$ the 5-dimensional Planck mass. In that case the self-consistency of RS model that includes bulk fields is in doubt. If, however, $k\lesssim M$, then an approximate self-consistency of the RS-type metric may still be satisfied.

\end{abstract}

\maketitle

\section{\label{sec:level1}INTRODUCTION}
The hierarchy between the Planck scale, $M_{Pl}\sim10^{19}$ GeV, and the standard model (SM) scale, $M_{SM}\sim1$ TeV, has been a long-standing problem in high energy physics. In the past decade, there have been two popular solutions to the hierarchy problems: the Arkani-Hamed-Dimopoulos-Dvali (ADD) model \cite{1} and the Randall-Sundrum (RS) model \cite{2}. Both models invoke extra dimensions and the brane-world scenario. The weakness of gravity is associated with the largeness of the extra dimension in the case of ADD, while in RS it is due to the exponential warpage of the extra dimension. In this paper we shall only focus our attention on the latter. In the RS model, two flat, parallel 3-branes are located at two fixed points on a $S^1/Z_2$ orbifold. The metric of RS reads:
\begin{eqnarray}
    ds^2=e^{-2\sigma(y)}\eta_{\mu\nu}dx^\mu dx^\nu+dy^2
    \label{metric1}.
\end{eqnarray}
where $R$ is the radius of the orbifold and $y$ is the extra dimension coordinate ranging from $-\pi R$ to $\pi R$ and $\sigma(y)=k|y|$. We adopt the convention $\eta_{\mu\nu}={\rm diag}(-1,1,1,1)$. The brane located at $y=0$ is called the hidden brane and that at $y=\pi R$ is the visible brane, on which the familiar SM fields reside. Under this construction, the hierarchy problem is naturally solved via the warp factor $e^{-k\pi R}$ along the extra dimension, generating a large hierarchy without requiring large extra dimensions. Specifically, with the choice of $kR\simeq12$, the TeV scale at the visible brane can be descended from the Planck scale at the hidden brane.

With the introduction of a finite separation between the two branes in the extra dimension, any field in the bulk should induce a Casimir energy. This provides a possible resolution to the smallness problem of the cosmological constant when connected with the observed dark energy \cite{3,4,5}. The Casimir effect in RS model generated by scalar fields has been studied in details by many authors \cite{6,Saharian1,Saharian2,Saharian3,Setare,8,scalarToms,Elizalde}. In \cite{6}, in particular, a detailed analysis of the \textsl{local} Casimir energy, or to be more specific, the vacuum expectation value (VEV) of stress-energy tensor, is given, in order to test the self-consistency of the model. It should be noted, as was pointed out by the authors of \cite{6}, however, that some conclusions made therein might not hold in a supersymmetric RS model, where the fermionic contribution is expected to exactly cancel its superpartner counterpart yet its local behavior has never been explicitly investigated. It is thus desirable to investigate the effect of the Casimir energy induced by the fermion field in the bulk. 

With regard to the fermionic Casimir effect in RS geometry, some authors obtained the result by multiplying a degeneracy factor to that of the scalar field \cite{9,BraneStability}. In such an approach, calculations are performed via the mode summation method. We note that while such a transcription from the scalar field Casimir energy to that of the spinor field in a flat spacetime geometry is clear, that for the curved spacetime is not as transparent. This is mainly because the reduction from the Dirac equation to the Klein-Gordon equation in flat spacetime is unambiguous, whereas the similar procedure in the curved spacetime would induce a curvature coupling term \cite{Dirac}:
\begin{equation}
\label{Diraccurve }
(-\nabla^2+m_\psi^2+\frac{1}{4}\hat{R})\psi=0,
\end{equation}
where $\psi$ is the fermion field, $\hat{R}$ is the Ricci scalar and $m_\psi$ is the mass of the fermion. On the other hand, for a scalar field $\phi$ in the curved spacetime, the equation of motion reads:
\begin{equation}
\label{scalarcurve }
(-\nabla^2+m_\phi^2+\xi \hat{R})\phi=0.
\end{equation}
The ambiguity arises because the coupling parameter $\xi$ between the scalar field and the curvature is generally not restricted to $\frac{1}{4}$, but as a free input to the theory. This motivates us to calculate the Casimir energy via the Green function method where the stress-energy tensor can be derived explicitly without resorting to the degeneracy factor.

We note that the vacuum energy arising from the fermion field in RS model has been fully discussed by Flachi \textsl{et al}. \cite{FerToms1}\cite{FerToms2} by computing the lowest order quantum corrections to the effective action. The finite temperature effect of the Casimir energy arising from both scalar field and spinor is discussed in \cite{Brevik}. These approaches can only provide the global properties of the vacuum energy and important local behavior might be overlooked. With the interest to fully understand the nature of the Casimir energy in the RS brane world, we find it desirable to directly compute the VEV for the stress-energy tensor, so as to compare it with that induced by the scalar field explicitly in the context of a SUSY system. 

\section{BASIC SETUP}
Let us first distinguish two sets of labels in our metric:
\begin{eqnarray}
    ds^2=g_{MN}dx^Mdx^N=e^{-2\sigma(y)}\eta_{\mu\nu}dx^\mu dx^\nu+dy^2
    \label{metric2},
\end{eqnarray}
where the capital Latin labels denote the (4+1)-dimensional quantities: $M=0,1,2,3,5$, so that $x^5=y$, whereas $\mu=0,1,..,3$. Then we write down the action for a Dirac spinor in the bulk with mass $m_{\Psi}=c\sigma'$ \cite{10}\cite{11}\cite{14}:
\begin{eqnarray}
   S=\int d^4x \int dy \sqrt{-g}\, \Big(i\bar{\Psi}\gamma^MD_M\Psi-m_\Psi\bar{\Psi}\Psi\Big)
    \label{action},
\end{eqnarray}
where $g={\rm det}(g_{MN})$. The gamma matrices, $\gamma^M$ are defined in curved spacetime as $\gamma^M=e_{\alpha}^{~M}\gamma^{\alpha}$, where $ e_{\alpha}^{~M}={\rm diag}(e^\sigma,e^\sigma,e^\sigma,e^\sigma,1)$ is the inverse vierbein and $\gamma^{\alpha}=(\gamma^a,i\gamma^5)$ are the gamma matrices in flat spacetime, and $D_M=\partial_M+\Gamma_M$ is the covariant derivative in curved spacetime. With the metric defined in (\ref{metric2}), we have
\begin{eqnarray}
\Gamma_\mu=-\frac{i}{2}\sigma'e_{\beta\mu}\gamma^5\gamma^\beta, \quad \Gamma_5=0
    \label{spinconnection}.
\end{eqnarray}
It should be noted that the equation of motion derived from (\ref{action}) with $m_\Psi=0$ is automatically conformally covariant under proper choice of boundary condition, without the need for an additional conformal coupling term\cite{12}.

Given the $Z_2$ transformation for Dirac fermion: $\Psi(-y)=\pm\gamma^5\Psi(y)$, it can be seen that $\bar{\Psi}\Psi$ is odd under $Z_2$ transformation. This implies that $m_{\Psi}$ must also be odd under $Z_2$ transformation in order to preserve the $Z_2$ symmetry of the Dirac equation. Therefore, $m_\Psi$ can be parametrized as \cite{10}
\begin{align}
m_\Psi=c\sigma'=ck\epsilon(y). \label{m}
\end{align}
where $\epsilon(y)$ is defined as being 1(-1) for positive(negative) $y$.

The stress-energy tensor related to the above action is given by
\begin{eqnarray}
T_{MN}=i\bar{\Psi}\gamma_{(M}D_{N)}\Psi-g_{MN}L
    \label{SET},
\end{eqnarray}
where the second term that involves the Lagrangian does not contribute to VEV by virtue of the equation of motion, and will be neglected in the following calculation.

\section{Computation of the Green Function}

In order to obtain the VEV of stress-energy tensor, we first calculate the Green function $G(x^M,x'^M)$ for the field, then express $\langle T_{MN}\rangle$ in terms of the Green function. The Green function by definition satisfies the field equation

\begin{eqnarray}
(i\gamma^MD_M-m_\Psi)G(x^M,x'^M)=\frac{1}{\sqrt{-g}}\delta(x^M-x'^M)
    \label{Greeneq}.
\end{eqnarray}
To eliminate the dependence on the coordinates of $x^\mu$, we perform a 4-dimensional Fourier transform on the Green function
\begin{align}
G(x^M,x'^M)&=\notag\\
&\int\frac{d^3\vec{p}}{(2\pi)^3}\int\frac{d\omega}{2\pi}e^{i\vec{p}\cdot(\vec{x}-\vec{x}')}e^{-i\omega(t-t')}G_p(y,y')
    \label{Fourier}.
\end{align}
Then (\ref{Greeneq}) becomes
\begin{eqnarray}
[-\gamma^\mu p_\mu-\gamma^5\partial_5-m_\Psi]\tilde{G}_p(y,y')=e^{2\sigma}\delta(y-y')I_{4\times4}
    \label{Greeneq2}.
\end{eqnarray}
where $\tilde{G}_p(y,y')\equiv e^{-2\sigma}G_p(y,y')$.

We further write the Green function as
\begin{eqnarray}
\tilde{G}_p(y,y')=\left(\begin{array}{cc}
                    f_1 & f_2 \\
                    g_1 & g_2
                  \end{array}\right)
\label{def:fg}.
\end{eqnarray}
where $f_1,f_2,g_1,g_2$ are $4\times4$ matrices. Put this back into (\ref{Greeneq2}), we arrive at the following 4 equations for the 4 elements of the Green function:
\begin{subequations}
\begin{align}
-\partial_5 f_1+ckf_1+e^\sigma\sigma^\alpha p_\alpha g_1=-e^{2\sigma}\delta(y-y')I_{2\times2}, \label{EOMa}
\\
e^\sigma\bar{\sigma}^\alpha p_\alpha f_1+\partial_5g_1+ckg_1=0, \label{EOMb}
\\
-\partial_5 f_2+ckf_2+e^\sigma\sigma^\alpha p_\alpha g_2=0, \label{EOMc}
\\
e^\sigma\bar{\sigma}^\alpha p_\alpha f_2+\partial_5g_2+ckg_2=-e^{2\sigma}\delta(y-y')I_{2\times2}. \label{EOMd}
\end{align}
\end{subequations}
If we neglect the Dirac delta functions on the right hand sides of (\ref{EOMa})$\sim$ (\ref{EOMd}), the solutions are\cite{10}
\begin{align}
f_i(y)=\frac{e^{\sigma/2}}{N_i}[J_{c-\frac{1}{2}}(mz)+b_iH^1_{c-\frac{1}{2}}(mz)],\label{solofEOM1}\\
g_i(y)=\frac{e^{\sigma/2}}{N_i}[J_{c+\frac{1}{2}}(mz)+d_iH^1_{c+\frac{1}{2}}(mz)].\label{solofEOM2}
\end{align}
where $m\equiv\sqrt{-p^2}$ is the Kaluza-Klein mass, $z\equiv e^{\sigma}/k$, $i=1,2$. $N_i$ and $b_i$ are the coefficients to be determined by normalization and by our choice of the boundary condition (BC). Generally, there are two distinct classes of BCs for Dirac fermion, the untwisted and the twisted BC\cite{10}\cite{13}. These BCs are derfined as follows.
\begin{subequations}
\begin{align}
\text{Untwisted:}\nonumber\\
 \Psi_L\mid_0=0&,&\Psi_L\mid_{\pi R}=0.\cr
                  (\partial_5+ck)\Psi_R\mid_0=0&,&(\partial_5+ck)\Psi_R\mid_{\pi R}=0
    \label{untBC}.\\[1mm]
 \text{Twisted:}\nonumber\\
 \Psi_L\mid_0=0&,&(\partial_5-ck)\Psi_L\mid_{\pi R}=0.\cr
 (\partial_5+ck)\Psi_R\mid_0=0&,&\Psi_R\mid_{\pi R}=0
    \label{tBC},
\end{align}
\end{subequations}
where $\Psi_{L,R}=[(1\mp\gamma^5)/2]\Psi$. We'll deal with both cases.

\subsection{\label{untwistedBC} The Untwisted BC}
For later convenience, we introduce the following four functions to represent different combinations of the special functions that appeared in (\ref{solofEOM1}) or (\ref{solofEOM2})
\begin{align}
\eta(y)\equiv e^{\frac{\sigma}{2}}[J_{c-\frac{1}{2}}(mz)+bH^1_{c-\frac{1}{2}}(mz)], \label{eta}\\
\acute{\eta}(y)\equiv e^{\frac{\sigma}{2}}[J_{c-\frac{1}{2}}(mz)+\acute{b}H^1_{c-\frac{1}{2}}(mz)], \label{acuteeta}\\
\lambda(y)\equiv-e^{\frac{\sigma}{2}}[J_{c+\frac{1}{2}}(mz)+bH^1_{c+\frac{1}{2}}(mz)], \label{lambda}\\
\acute{\lambda}(y)\equiv-e^{\frac{\sigma}{2}}[J_{c+\frac{1}{2}}(mz)+\acute{b}H^1_{c+\frac{1}{2}}(mz)], \label{acutelambda}
\end{align}
where
\begin{align}
b=-\frac{J_{c-\frac{1}{2}}(\frac{m}{k})}{H^1_{c-\frac{1}{2}}(\frac{m}{k})}, \label{defb}\\
\acute{b}=-\frac{J_{c-\frac{1}{2}}(\frac{me^{\pi kR}}{k})}{H^1_{c-\frac{1}{2}}(\frac{me^{\pi kR}}{k})}. \label{defacuteb}
\end{align}
In terms of these new functions, the untwisted BCs become
\begin{align}
 \eta\mid_0=0&,&\acute{\eta}\mid_{\pi R}=0.\cr
                  (\partial_5+ck)\lambda\mid_0=0&,&(\partial_5+ck)\acute{\lambda}\mid_{\pi R}=0
    \label{etalambdaBC}.
\end{align}
We first deal with $f_1$ and $g_1$. Under the above definitions, $f_1$ and $g_1$ can be expressed as
\begin{align}
f_1(y,y') = \left\{\begin{array}{ll}
                 \left(\begin{array}{cc}
                   \alpha & \gamma \\
                   \bar{\alpha} & \bar{\gamma}
                 \end{array}\right)\eta(y)
                 , & \mbox{if $y<y'$} \\[5mm]
                 \left(\begin{array}{cc}
                   \acute{\alpha} & \acute{\gamma} \\
                   \acute{\bar{\alpha}} & \acute{\bar{\gamma}}
                 \end{array}\right)\acute{\eta}(y), & \mbox{if $y>y'$} \\
                \end{array}\right .\label{f1} \\[5mm]
g_1(y,y') = \left\{\begin{array}{ll}
                 \left(\begin{array}{cc}
                   \beta & \delta \\
                   \bar{\beta} & \bar{\delta}
                 \end{array}\right)\lambda(y)
                 , & \mbox{if $y<y'$} \\[5mm]
                 \left(\begin{array}{cc}
                   \acute{\beta} & \acute{\delta} \\
                   \acute{\bar{\beta}} & \acute{\bar{\delta}}
                 \end{array}\right)\acute{\lambda}(y), & \mbox{if $y>y'$} \\
                \end{array}\right . \label{g1}
\end{align}
All the parameters in the above matrices are determined by (\ref{EOMa}) and (\ref{EOMb}), from which we get
\begin{align}
&g_1\mid_{y=y'^+}-g_1\mid_{y=y'^-}=0, \label{constraint1}\\
&f_1\mid_{y=y'^+}-f_1\mid_{y=y'^-}=e^{2\sigma(y')}I_{2\times2}, \label{constraint2}\\
&-\partial_5f_1+ckf_1+e^\sigma\sigma^\alpha p_\alpha g_1=0. \label{constraint3}
\end{align}

These together constitute 16 linear equations for 16 unknowns. Here we solved it by the help of \verb"Mathematica 7.0":
\begin{align}
&f_1(y,y') = \left\{\begin{array}{ll}
                 -I_{2\times2}e^{2\sigma(y')}\frac{\acute{\lambda}(y')}{S}\eta(y), & \mbox{if $y<y'$} \\[1mm]
                 -I_{2\times2}e^{2\sigma(y')}\frac{\lambda(y')}{S}\acute{\eta}(y), & \mbox{if $y>y'$} \\
                \end{array}\right .\label{f1sol} \\[2mm]
&g_1(y,y') = \left\{\begin{array}{ll}
                 -\bar{\sigma}^\alpha p_\alpha e^{2\sigma(y')}\frac{\acute{\lambda}(y')}{mS}\lambda(y), & \mbox{if $y<y'$} \\[1mm]
                 -\bar{\sigma}^\alpha p_\alpha e^{2\sigma(y')}\frac{\lambda(y')}{mS}\acute{\lambda}(y), & \mbox{if $y>y'$} \\
                \end{array}\right .\label{g1sol}
\end{align}
where
\begin{align}
S\equiv\acute{\lambda}\eta-\lambda\acute{\eta}\mid_{y'}. \label{S}
\end{align}
The remaining two elements, $f_2$ and $g_2$, can be obtained in a similar way

\begin{align}
&f_2(y,y') = \left\{\begin{array}{ll}
                 -\sigma^\alpha p_\alpha e^{2\sigma(y')}\frac{\acute{\eta}(y')}{mS}\eta(y), & \mbox{if $y<y'$} \\[1mm]
                 -\sigma^\alpha p_\alpha e^{2\sigma(y')}\frac{\eta(y')}{mS}\acute{\eta}(y), & \mbox{if $y>y'$} \\
                \end{array}\right .\label{f2sol} \\[2mm]
&g_2(y,y') = \left\{\begin{array}{ll}
                 -I_{2\times2}e^{2\sigma(y')}\frac{\acute{\eta}(y')}{S}\lambda(y), & \mbox{if $y<y'$} \\[1mm]
                 -I_{2\times2}e^{2\sigma(y')}\frac{\eta(y')}{S}\acute{\lambda}(y), & \mbox{if $y>y'$} \\
               \end{array}\right .\label{g2sol}
\end{align}
With (\ref{f1sol}), (\ref{g1sol}), (\ref{f2sol}), (\ref{g2sol}), we finish our calculation for the Green function subjected to the untwisted BC.

\subsection{\label{untwistedBC} The Twisted BC}
In order to fit the twisted BC (\ref{tBC}), we define additional two functions
\begin{align}
\tilde{\eta}(y)\equiv-e^{\frac{\sigma}{2}}[J_{c+\frac{1}{2}}(mz)+\tilde{b}H^1_{c+\frac{1}{2}}(mz)], \label{tildeeta}\\
\tilde{\lambda}(y)\equiv e^{\frac{\sigma}{2}}[J_{c-\frac{1}{2}}(mz)+\tilde{b}H^1_{c-\frac{1}{2}}(mz)], \label{tildelambda}
\end{align}
where
\begin{align}
\tilde{b}=-\frac{J_{c+\frac{1}{2}}(\frac{me^{\pi kR}}{k})}{H^1_{c+\frac{1}{2}}(\frac{me^{\pi kR}}{k})}, \label{deftildeb}
\end{align}
so that
\begin{align}
\tilde{\eta}\mid_{\pi R}=0, ~~(\partial_5-ck)\tilde{\lambda}\mid_{\pi R}=0. \label{tildebBC}
\end{align}

If we rewrite (\ref{f1}) and (\ref{g1}) with the replacements $\acute{\eta}\rightarrow\tilde{\lambda}$ and $\acute{\lambda}\rightarrow\tilde{\eta}$, the remaining calculations are completely identical to that of the untwisted case. The results are just the same as (\ref{f1sol}), (\ref{g1sol}), (\ref{f2sol}) and (\ref{g2sol}) with the above substitution.

\section{Stress-Energy Tensor}
Now well-equipped with the exact form of the Green functions for both untwisted and twisted BCs, namely, (\ref{f1sol})$\sim$ (\ref{g2sol}), together with the definition (\ref{def:fg}), we are finally at a position to calculate the VEV of the stress-energy tensor. First, we make the identification
\begin{align}
iG(x^M,x'^M))=\langle\Psi(x^M)\bar{\Psi}(x'^M)\rangle. \label{id}
\end{align}
With this identification, we can replace any VEV involving the quadratic terms of the field by the Green function. We begin with the VEV of $T_{00}$ in (\ref{SET}),
\begin{align}
\langle T_{00}\rangle&=ie_{00}\langle\bar{\Psi}\gamma^0D_0\Psi\rangle\notag\\
&=-ie^{-\sigma}[i\partial_0\mbox{Tr}(\gamma^0G)]_{x^M=x'^M}. \label{T00}
\end{align}
In the second equality we have neglected the spin connection term $\Gamma_0$ in the covariant derivative $D_0$, since it turns out to be a term independent of the mode, and thus can be omitted in the renormalization. Let us first deal with the untwisted BC. We perform a 4-dimensional Fourier transform on both sides of (\ref{T00}) and use (\ref{def:fg}) to obtain
\begin{align}
\langle t_{00}\rangle&=-i\omega e^\sigma {\rm Tr}(g_1+f_2)|_{y=y'}\notag\\
&=-2ie^{3\sigma}\frac{\omega^2}{m}\frac{\acute{\lambda}\lambda+\acute{\eta}\eta}{\acute{\lambda}\eta-\lambda\acute{\eta}}\quad \mbox{(Untwisted BC)}, \label{t00}
\end{align}
where $t_{00}$ is the 4-dimensional Fourier transformation of $T_{00}$. In a similar fashion, we obtain the other entries of the stress-energy and also those for the twisted BC as follows.
\begin{align}
&\langle T_{\mu\nu}\rangle =  \notag\\
&\left\{\begin{array}{ll}
                 -2i\delta_{\mu\nu}e^{3\sigma}\int\frac{d^3\vec{p}}{(2\pi)^3}\int\frac{d\omega}{2\pi}\frac{(p^\mu)^2}{m}\frac{\acute{\lambda}\lambda+\acute{\eta}\eta}{\acute{\lambda}\eta-\lambda\acute{\eta}}, \quad\mbox{untwisted}\\
                 -2i\delta_{\mu\nu}e^{3\sigma}\int\frac{d^3\vec{p}}{(2\pi)^3}\int\frac{d\omega}{2\pi}\frac{(p^\mu)^2}{m}\frac{\tilde{\eta}\lambda+\tilde{\lambda}\eta}{\tilde{\eta}\eta-\lambda\tilde{\lambda}}, \quad\mbox{twisted}\\
                 \end{array}\right .
\label{Tmunu}\\
&\langle T_{yy}\rangle = \notag\\
&\left\{\begin{array}{ll}
                 -ie^{4\sigma}\int\frac{d^3\vec{p}}{(2\pi)^3}\int\frac{d\omega}{2\pi}\,\partial_5(\frac{\acute{\lambda}\lambda+\acute{\eta}\eta}{\acute{\lambda}\eta-\lambda\acute{\eta}})\mid_{y=y'}, \mbox{untwisted}\\
                 -ie^{4\sigma}\int\frac{d^3\vec{p}}{(2\pi)^3}\int\frac{d\omega}{2\pi}\,\partial_5(\frac{\tilde{\eta}\lambda+\tilde{\lambda}\eta}{\tilde{\eta}\eta-\lambda\tilde{\lambda}})\mid_{y=y'}, \quad\mbox{twisted}\\
                 \end{array}\right .
\label{Tyy}
\end{align}
where $\delta_{\mu\nu}={\rm diag}(1,1,1,1)$. This is our final expression for the stress-energy tensor. The integral cannot be expressed analytically for a general $c$, and suitable regularization must be implemented. For the massless case, however, the results can be written in a concise way.

\section{Massless Dirac Fermion}
Now consider the simplest case: $c=0$, in which all $\eta$ and $\lambda$ functions reduce to basic trigonometric functions, and (\ref{Tmunu}) and (\ref{Tyy}) become
\begin{align}
&\langle T_{\mu\nu}\rangle = \notag\\
                 &\left\{\begin{array}{ll}
                 -\!&\!2i\delta_{\mu\nu}e^{3\sigma} \!\! \int \!\! \frac{d^3\vec{p}}{(2\pi)^3}\int \!\! \frac{d\omega}{2\pi}\frac{(p^\mu)^2}{m}\,\mbox{cot}[\frac{m}{k}(e^{\pi kR}-1)], \!\mbox{untwisted}\\
                 \!&\!2i\delta_{\mu\nu}e^{3\sigma} \!\! \int \!\! \frac{d^3\vec{p}}{(2\pi)^3}\int \!\! \frac{d\omega}{2\pi}\frac{(p^\mu)^2}{m}\,\mbox{tan}[\frac{m}{k}(e^{\pi kR}-1)], \mbox{twisted}\\
                 \end{array}\right .
\label{c0Tmunu}\\
&\langle T_{yy}\rangle =  \notag\\
                 &\left\{\begin{array}{ll}
                 -&2ie^{5\sigma}\int\frac{d^3\vec{p}}{(2\pi)^3}\int\frac{d\omega}{2\pi}\,m\,\mbox{cot}[\frac{m}{k}(e^{\pi kR}-1)], \,\mbox{untwisted}\\
                 &2ie^{5\sigma}\int\frac{d^3\vec{p}}{(2\pi)^3}\int\frac{d\omega}{2\pi}\,m\,\mbox{tan}[\frac{m}{k}(e^{\pi kR}-1)], \,\mbox{twisted}\\
                 \end{array}\right .
\label{c0Tyy}
\end{align}
The integrations are best done by performing a Wick rotation
\begin{align}
\omega\rightarrow ip^4,\quad m=\sqrt{\omega^2-\vec{p}^2}\rightarrow ip. \label{Wick}
\end{align}
Thus the stress-energy tensor is given by
\begin{align}
&\langle T_{\mu\nu}\rangle = \notag\\
                 &\left\{\begin{array}{ll}
                 -2\eta_{\mu\nu}e^{3\sigma}\int\frac{d^4p}{(2\pi)^4}\frac{(p^\mu)^2}{p}\,\mbox{coth}[\frac{p}{k}(e^{\pi kR}-1)], \quad\mbox{untwisted}\\
                 -2\eta_{\mu\nu}e^{3\sigma}\int\frac{d^4p}{(2\pi)^4}\frac{(p^\mu)^2}{p}\,\mbox{tanh}[\frac{p}{k}(e^{\pi kR}-1)], \quad\mbox{twisted}\\
                 \end{array}\right .
\label{Wickc0Tmunu}\\
&\langle T_{yy}\rangle =  \notag\\
                 &\left\{\begin{array}{ll}
                 2e^{5\sigma}\int\frac{d^4p}{(2\pi)^4}\,p\,\mbox{coth}[\frac{p}{k}(e^{\pi kR}-1)], \quad\mbox{untwisted}\\
                 2e^{5\sigma}\int\frac{d^4p}{(2\pi)^4}\,p\,\mbox{tanh}[\frac{p}{k}(e^{\pi kR}-1)], \quad\mbox{twisted}\\
                 \end{array}\right .
\label{Wickc0Tyy}
\end{align}
The regularization procedure for the integral is conventional, where one subtracts $1$ from the coth or tanh function. The results read (for a more detailed calculation, see, for example, \cite{6} or \cite{15})
\begin{align}
&\langle T_{\mu\nu}\rangle_{\rm ren}=\notag\\
                 &\left\{\begin{array}{ll}
                 -e^{3\sigma}\eta_{\mu\nu}2^{-3}\pi^{-\frac{5}{2}}a^{-5}\Gamma(\frac{5}{2})\zeta(5)&, \,\mbox{untwisted}\\
                 \frac{15}{16}e^{3\sigma}\eta_{\mu\nu}2^{-3}\pi^{-\frac{5}{2}}a^{-5}\Gamma(\frac{5}{2})\zeta(5)&, \,\mbox{twisted}\\
                 \end{array}\right .
\label{result:Tmunu}\\
&\langle T_{yy}\rangle_{\rm ren}=\notag\\
             &\left\{\begin{array}{ll}
                 \frac{1}{2}e^{5\sigma}\pi^{-\frac{5}{2}}a^{-5}\Gamma(\frac{5}{2})\zeta(5)&, \,\mbox{untwisted}\\
                -\frac{15}{32}e^{5\sigma}\pi^{-\frac{5}{2}}a^{-5}\Gamma(\frac{5}{2})\zeta(5)&, \,\mbox{twisted}\\
                 \end{array}\right .
\label{result:Tyy}
\end{align}
where $a=(e^{\pi kR}-1)/k$. Compare with the VEV of the stress-energy tensor for scalar fields given in \cite{6}, we conclude that
\begin{eqnarray}
\langle T_{MN}\rangle_{\rm Dirac, untwisted}=-4\langle T_{MN}\rangle_{\rm real~scalar}\\
\langle T_{MN}\rangle_{\rm Dirac, twisted}=\frac{15}{4}\langle T_{MN}\rangle_{\rm real~scalar}\label{comparison}
\end{eqnarray}
The factor 4, as indicated in \cite{9}, accounts for the difference of the degrees of freedom between the Dirac spinor and the real scalar fields. The sign difference originates from the distinct natures of fermions and bosons. As for the factor $15/16$ between the untwisted and the twisted results, it results from the difference in regularization between the coth and the sinh functions.

\section{MASSIVE DIRAC FERMION}
To extract useful information from the VEV of stress-energy tensor for a general mass $c$, we need to appeal to the numerical method. In this section, we will only focus on the numerical integration of the 00 component of $\langle T_{MN}\rangle$ for the \textsl{untwisted BC} (the general features, including the power and the signs of surface divergences, are the same for the twisted BC), namely, (\ref{t00}). For later convenience, we denote
\begin{equation}
\label{F}
F(p,y;R)\equiv \frac{\acute{\lambda}\lambda+\acute{\eta}\eta}{\acute{\lambda}\eta-\lambda\acute{\eta}}.
\end{equation}
Recall that $R$ is the radius of the RS geometry. So our goal is to regularize the integral
\begin{equation}
\label{goal}
\langle T_{00}\rangle = iC \int_0^\infty dp~p^4 F(ip,y;R),
\end{equation}
with
\begin{equation}
\label{C }
C\equiv \frac{\pi^{-\frac{5}{2}}}{4}\frac{\Gamma(\frac{3}{2})}{\Gamma(3)}e^{3\sigma(y)}.
\end{equation}
Note that the Wick rotation has been performed in the above expression.

To regularize the integral, we subtract the same quantity, namely, 1, as the massless case from the integrand $iF(ip,y;R)$. In other word, since $iF(ip,y;R)$ approches unity as $p\rightarrow \infty$, we are in fact subtracting its ultraviolet limit\cite{Milton}.
\begin{equation}
\label{rengoal}
\langle T_{00}\rangle_{\rm ren} = C \int_0^\infty dp~p^4 [iF(ip,y;R)-1]
\end{equation}

To facilitate the computation, we expand $iF(ip,y;R)$ to the first order in $c$. Since $c$ represents the fermion mass in units of Planck mass, it is expected to be very small for a physical particle, thus ensuring the validity of our approximation. In addition, to make the computer run more effectively, we set $k=R=1$ as in \cite{6}, although the more ``realistic" value would be $kR\simeq12$ with $k\simeq M_{Pl}$.

\begin{figure}[t] %  figure placement: here, top, bottom, or page
\centering
   \includegraphics[width=8cm]{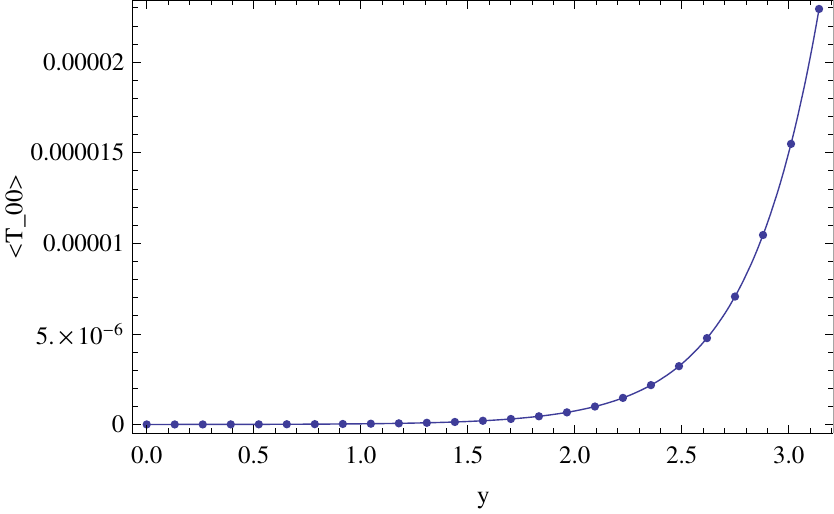} 
   \caption{\label{fig:1}The massless ($c$=0), untwisted BC case for $\langle T_{00}\rangle_{\rm ren}$ in units of $k^5$. The branes are located at $y=0$ and $y=\pi$ and $kR=1$. Note that the numerical value (points) agrees perfectly with the exact solution.}
\end{figure}

\begin{figure}[t] %  figure placement: here, top, bottom, or page
\centering
   \includegraphics[width=8cm]{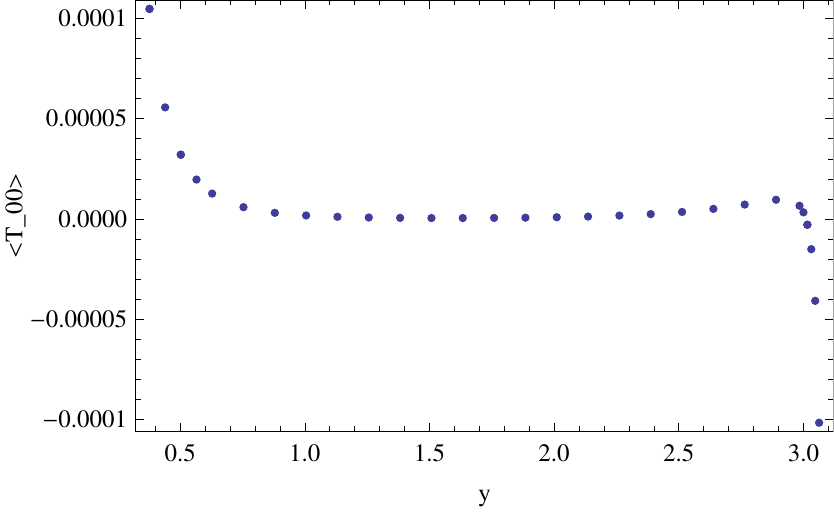} 
   \caption{\label{fig:2}The massive ($c=10^{-3}$), untwisted BC case for $\langle T_{00}\rangle_{\rm ren}$ in units of $k^5$. The energy density diverges at both branes, but with opposite signs.}
\end{figure}

The results are given in Fig.~\ref{fig:1} and Fig.~\ref{fig:2}. It can be seen that the energy density diverges to infinity near both branes, but with opposite signs. We'll discuss this feature in the following subsections.

\subsection{Surface Divergence}
In Ref.\cite{Candelas}, the authors pointed out that the surface divergence is mainly contributed by the high wave number (momentum) modes in the Fourier transformation of $\langle T_{MN} \rangle$. Based on the same philosophy, we expand the integrand of (\ref{rengoal}) asymptotically in powers of $p^{-n}$ and retain only the $p^{-1}$ term, which dominates the integrand for large $p$, i.e.,
\begin{align}
\label{SD1 }
&\langle T_{00} \rangle_{\rm ren} \simeq C\int_0^\infty  dp~p^4\Big[\mbox{coth}(\frac{\alpha-1}{k})p -1\Big]\notag\\
&+C\int_0^\infty dp~p^4 \frac{c}{2z\alpha p}\Big[\mbox{csch}(\frac{\alpha-1}{k})p\Big]^2\notag\\
&\times\Big\{z(\alpha-1)+\alpha\mbox{cosh}[2p(\frac{\alpha}{k}-z)]-\alpha\mbox{cosh}[2p(z-\frac{1}{k})]\Big\},
\end{align}
where $\alpha \equiv e^{\pi kR}$, and recall that $z=e^\sigma/k$. The first term is just the result for massless fermion while the second term corresponds to the correction due to the fermion mass. Note that the mass correction term is proportional to $c$ by virtue of the small mass expansion to the first order.

The expression remains complicated. However, since only the high wave number behavior concerns us, we isolate the dominating term by taking the large $p$ limit again and find
\begin{equation}
\label{SD2}
\langle T_{00} \rangle_{\rm SD} \sim C\int^\infty dp~\frac{2cp^3}{z}\Big[e^{-2p(z-\frac{1}{k})}-e^{-2p(\frac{\alpha}{k}-z)}\Big].
\end{equation}
(The subscipt SD stands for the surface divergent term). Since $z\equiv \frac{e^{ky}}{k}$ ranges from $\frac{1}{k}$ to $\alpha=\frac{e^{\pi kR}}{k}$, we find the near-brane behaviors to be
\begin{align}
\label{SD3}
&\langle T_{00} \rangle_{\rm ren}|_{y\rightarrow 0}\propto \frac{ce^{3ky}}{y^4},\\\label{SD4}
&\langle T_{00} \rangle_{\rm ren}|_{y\rightarrow \pi R}\propto \frac{-ce^{3ky}}{(y-\pi R)^4}.
\end{align}
We observe that while the energy density diverges to minus infinity at the visible brane, that near the hidden brane approaches to plus infinity. This surface divergence agrees qualitatively with our numerical result shown in Fig.~\ref{fig:2}. It should be noted that the divergence only disappears in the limit of $c=0$, the massless and conformally symmetric case, which is consistent with our previous result.

At the first glance, the asymmetry aspect of the surface divergence seems curious. One might blame the asymmetry to the nature of the RS geometry. However, this is not the case; the fermionic Casimir energy density in the flat case $k=0$ also possesses such asymmetric surfact divergence.

\subsection{The $k=0$ Case}

In the $k=0$ flat case with untwisted BC, where we set the two branes to be located at $y=a_0$ and $y=a_1$, the (\ref{eta})$\sim$(\ref{acutelambda}) should be replaced by
\begin{align}
&\eta(y)=\mbox{sin}K(y-a_0), \label{eta0}\\
&\acute\eta(y)=\mbox{sin}K(y-a_1), \label{acuteeta0}\\
&\lambda(y)=\mbox{cos}K(y-a_0)-\frac{m_{\Psi}}{K}\mbox{sin}K(y-a_0), \label{lambda0}\\
&\acute\lambda(y)=\mbox{cos}K(y-a_1)-\frac{m_{\Psi}}{K}\mbox{sin}K(y-a_1), \label{acutelambda0}
\end{align}
where $K\equiv \sqrt{m^2-m_{\Psi}^2}$ and $m$ is the KK mass defined as before. Putting the above functions into (\ref{F}) and (\ref{goal}), performing a Wick rotation upon $m$, we find a neat expression for the $k=0$ case.
\begin{equation}
\langle T_{00}\rangle_{k=0} = C(k=0) \int_0^\infty dp~p^4 \Big( T^{(0)}+T^{(1)}+T^{(2)}\Big),
\end{equation}
where
\begin{align}
&T^{(0)}\equiv \mbox{coth}[\kappa (a_1-a_0)y],\\
&T^{(1)}\equiv -\frac{m_\Psi}{\kappa}\frac{\mbox{sinh} [\kappa(2y-a_1-a_0)]}{\mbox{sinh}[\kappa(a_1-a_0)]},\\
&T^{(2)}\equiv \Big(\frac{m_\Psi}{\kappa}\Big)^2\frac{\mbox{sinh}[\kappa(y-a_0)]\mbox{sinh}[\kappa(y-a_1)]}{\mbox{sinh}[\kappa(a_1-a_0)]},
\end{align}
with $\kappa\equiv \sqrt{p^2+m_{\Psi}^2}$, the Wick rotated variable of $K$. The superscript $(i)$ stands for the $i$th order term (if not counting the $\kappa$ dependence) of the fermion mass $m_\Psi$. Taking the ultraviolet limit, we find that the latter two terms contribute to the surface divergence feature:
\begin{align}
\langle T_{00}\rangle_{k=0, {\rm SD}}\sim \int ^{\infty}dp~p^4\frac{m_\Psi}{\kappa}\Big[e^{-2\kappa(y-a_0)}-e^{-2\kappa(a_1-y)}\Big]\notag\\  
+\int ^{\infty}dp~\frac{p^4}{2}\Big(\frac{m_\Psi}{\kappa}\Big)^2\Big[e^{-2\kappa(y-a_0)}+e^{-2\kappa(a_1-y)}-1\Big]. \label{SD0}
\end{align}
The first line comes from $T^{(1)}$, the first order mass term, and the second line comes from $T^{(2)}$. The minus one in last parenthesis is independent of $a_1$ and $a_0$, thus could be eliminated in the regularization. Compare the leading order term in (\ref{SD0}) with (\ref{SD2}), we note that the surface divergence features are basically the same, so the asymmetry aspect doesn't seem to originate from the nature of RS geometry, but a generic feature for a 5D fermionic Casimir energy density. In addition, we note an interesting fact in (\ref{SD0}): the surface divergent term is antisymmetric with respect to the two branes for the first order mass term, and symmetric for the second order mass term.

In fact, the asymmetry aspect of the energy density originates from the parity transformation $y\rightarrow -y$. Under the parity, the fermion field mass changes sign,
\begin{equation}
m_{\Psi}\rightarrow -m_{\Psi}.
\end{equation}
As a result, all the terms proportional to the \textsl{odd} power of mass will change sign, while those proportional to the \textsl{even} power remain the same under the parity transformation. Thus the terms in the energy density will be antisymmetric for odd power of $m_\Psi$, and symmetric for even power ones. This is in accordance with our result (\ref{SD0}) and (\ref{SD2}). In other words, \textsl{the sign of the mass will determine a specific direction in the fifth dimension and thus break the exchange symmetry of the two branes.} It is also worth noting that such asymmetry aspect doesn't arise in the scalar field \cite{Milton,6}, since the mass of the scalar field maintains its sign under parity transformation.

With regard to the non-integrable divergence of the stress-energy tensor, the problem was solved by Kennedy, Critchley, and Dowker \cite{Kennedy} for a scalar field in a static spacetime, and further elaborated by Romeo and Saharian \cite{Romeo1,Romeo2}, and put in a broader context by Fulling 
 \cite{Fulling}. The resolution lies in the renormalization of the bare surface gravitational action, which induces a delta function and that cancels exactly with the surface divergence. 

Since the surface divergence should be renormalized into the surface action terms, we do \textsl{not} expect the powers of surface divergence for scalar field and Dirac spinor field to be the same, so that they could cancel each other in a supersymmetric theory. In fact, while the stress-energy tensor for a massive scalar field diverges cubically near the surface (see (2.39a) in \cite{Milton}),  that for a massive Dirac spinor field diverges quartically as in (\ref{SD3}) and (\ref{SD4}), and that for a minimally-coupled massless scalar field diverges as inverse fifth power of the distance (see \cite{6}). The differences could be understood by simple power-counting analysis. 

Generally speaking, a surface divergence term in flat (4+1)-dimension takes the form
\begin{equation}
\langle T_{00} \rangle\sim \int dp~p^4~\frac{\Omega}{p^\gamma}u(py),
\end{equation}
with $\Omega$ the quantity, the mass for example, that brings about the surface divergence, and $\gamma$ the mass dimension of it. Since the Casimir energy density has mass dimension 5 in (4+1)-dimension, $u(py)$ is dimensionless. Consider a minimally-coupled massless scalar field, the quantity signaling the surface divergence is the dimensionless coupling constant, which means that $\Omega$ is dimensionless and $\gamma=0$, so the integration yields a $y^{-5}$ surface divergence. On the other hand, $\Omega=ck=m_\Psi$ for a massive Dirac spinor field, so $\gamma=1$ and the surface divergence is of inverse fourth power of the distance from the surface. As for the massive scalar case, since the mass of a scalar field generally appears in power of $m_\phi^2$, $\Omega=m_\phi^2$ and $\gamma=2$ for a massive scalar field, hence the surface divergent term behaves as $y^{-3}$.

\section{DISCUSSIONS and conclusions}
In flat spacetime, the one-loop correction to the effective action, and therefore the Casimir energy, of the massless Dirac spinor field can be obtained simply by multiplying a suitable degeneracy factor to that of the scalar field. In the case of curved spacetime, however, the above statement is not valid in general. In this paper, we provide a complete and straightforward derivation, using the Green function approach, of these degeneracy factors, which are $-4$ and $15/4$ for the untwisted and twisted BC, respectively, as expected.
For a massive fermion, the Casimir energy density is plotted as a function of the position $y$ in Fig.\ref{fig:2}. The asymmetry aspect of the surface divergence comes from the nature of the parity transformation in the fifth dimension, which can also be found in the flat case.

To cure the surface divergence, an approach similar to that of \cite{Kennedy} must be implemented where the surface terms are included. Under proper renormalization of the surface action term, the SUSY cancellation is not threatened by the power difference of the surface divergence between the scalar and the spinor fields. On the other hand, this infinity might be originated from the unphysical nature of the boundary condition. This may suggest the necessity of including the finite thickness (of string length) of the 3-brane in the treatment. In fact, the finite thickness effect of a de Sitter brane has been discussed in \cite{dS1,dS2,dS3}. Similar situation happens in the perfect conductor boundary condition for electromagnetic field \cite{Candelas,Balian1,Balian2}, in which a strictly zero thickness boundary gives rise to a non-integrable divergence. For an imperfect conductor, such as a dielectric material, or a conductor with finite thickness, waves of sufficiently high frequency would penetrate into the material so that the precise location of the boundary would lose its meaning. Therefore, in reality, the divergence does not appear since the expression of the integrand is not universally applicable for all wave numbers. The finite thickness of the 3-brane might provide a similar remedy to this divergence problem.

Last, but not the least, there is the issue of self-consistency for the RS metric to retain its solution under the quantum corrected Einstein equations. As Knapman and Toms \cite{6} commented in the case of the massless, conformally-coupled scalar field, different components of the stress energy-momentum tensor would in general induce different corrections to $k$, if indeed an RS-type solution can be found. It was argued \cite{6} that such correction is exponentially small and therefore the RS solution remains approximately self-consistent. We found, however, that in both cases of the massless Dirac spinor field and the conformally-coupled scalar field, the VEV of the stress-energy tensor takes the form (see (\ref{result:Tmunu}) and (\ref{result:Tyy})):
\begin{align}
\langle T_{\mu\nu} \rangle= \mathcal{O}(1)\times e^{3k|y|-5k\pi R}k^5,\\
\langle T_{yy} \rangle= \mathcal{O}(1)\times e^{5k|y|-5k\pi R}k^5,
\end{align}
where $\mathcal{O}(1)$ is some quantities of the order unity. On the other hand, the cosmological constant term in the Einstein equation is $\Lambda g_{MN}\sim -M^3k^2 g_{MN}$ \cite{2}, where $g_{MN}$ is given by (\ref{metric2}), so the ratio of stress-energy tensor to the classical one is about 
\begin{equation}
(k/M)^3e^{5k(|y|-\pi R)}.
\end{equation}
 If $k\simeq M \simeq M_{Pl}$, due to the requirement of no hierarchy problem, then the stress-energy tensor is \textsl{comparable} to the cosmological constant term in the region where $y\rightarrow \pi R$, even though it is small with respect to $k^5$ in most regions. As a result, while the quantum correction of the metric is exponentially small at the UV-brane, that at the visible brane is of the same order as that of the classical Einstein equation with RS geometry. Thus we cannot treat the VEV of the stress-energy tensor as a perturbation to the semi-classical Einstein equation, and the metric of the form (\ref{metric2}) will not be a solution to the quantum corrected Einstein equation, not even in the approximate sense as argued in \cite{6}. The self-consistency of the RS model that includes bulk fields is therefore in doubt. 

However \footnote{We are indebted to the referee for this remark.}, if $k$ is slightly smaller than $M$ as commented in \cite{2}, then the suppression $(k/M)^3$ would be rather significant even near the visible brane. Therefore, an \textsl{approximate} self-consistency of the RS-type metric may still be satisfied.

\begin{acknowledgments}
We thank L. P. Teo and P. M. Ho for their useful suggestions and comments on the subject. We are also grateful to K. Y. Su, Y. D. Huang and C. I. Chiang for interesting and encouraging discussions. This research is supported by Taiwan National Science Council under Project No. NSC 97-2112-M-002-026-MY3, by Taiwan's National Center for Theoretical Sciences (NCTS), and by US Department of Energy under Contract No. DE-AC03-76SF00515.
\end{acknowledgments}

\end{document}